\overfullrule=0pt
\input harvmac
\def\a{{\alpha}}

\def\l{{\lambda}}
\def\lh{{\widehat\lambda}}
\def\b{{\beta}}

\def\g{{\gamma}}

\def\d{{\delta}}

\def\N{{\nabla}}
\def\Nb{{\overline\nabla}}

\def\O{{\Omega}}

\def\o{{\omega}}
\def\oh{{\widehat\omega}}
\def\half{{1\over 2}}
\def\p{{\partial}}
\def\pb{{\overline\partial}}
\def\t{{\theta}}

\def\oh{{\widehat\o}}
\def\L{{\Lambda}}

\def\Pib{{\overline\Pi}}

\def\Gb{\overline\Gamma}

\baselineskip12pt

\Title{ \vbox{\baselineskip12pt
}}
{\vbox{\centerline
{Simplified Pure Spinor $b$ Ghost in a  }
\bigskip
\centerline{Curved Heterotic Superstring Background  }
}}
\smallskip
\centerline{Nathan Berkovits\foot{e-mail: nberkovi@ift.unesp.br}, }
\smallskip
\centerline{\it ICTP South American Institute for Fundamental Research}
\centerline{\it Instituto de F\'{\i}sica Te\'orica, UNESP - Univ. Estadual Paulista}
\centerline{\it Rua Dr. Bento T. Ferraz 271, 01140-070, S\~ao Paulo, SP, Brasil} 
\bigskip
\centerline{Osvaldo Chandia\foot{e-mail: ochandiaq@gmail.com}, }
\smallskip
\centerline{\it Departamento de Ciencias, Facultad de Artes Liberales, Universidad Adolfo Ib\'a\~nez}
\centerline{\it Facultad de Ingenier\'{\i}a y Ciencias, Universidad Adolfo Ib\'a\~nez}
\centerline{\it Diagonal Las Torres 2640, Pe\~nalol\'en, Santiago, Chile}

\bigskip

\noindent
Using the RNS-like fermionic vector variables introduced in arXiv:1305.0693, the pure spinor $b$ ghost in a curved heterotic superstring background is easily constructed. This construction simplifies and completes the $b$ ghost construction in a curved background of 
arXiv:1311.7012.

\Date{March 2014}


\lref\BerkovitsFE{
  N.~Berkovits,
  ``Super Poincare covariant quantization of the superstring,''
JHEP {\bf 0004}, 018 (2000).
[hep-th/0001035].
}

\lref\BerkovitsBT{
  N.~Berkovits,
  ``Pure spinor formalism as an N=2 topological string,''
JHEP {\bf 0510}, 089 (2005).
[hep-th/0509120].
}

\lref\BerkovitsPLA{
  N.~Berkovits,
  ``Dynamical twisting and the b ghost in the pure spinor formalism,''
JHEP {\bf 1306}, 091 (2013).
[arXiv:1305.0693 [hep-th]].
}

\lref\ChandiaIMA{
  O.~Chandia,
  ``The Non-minimal Heterotic Pure Spinor String in a Curved Background,''
[arXiv:1311.7012 [hep-th]].
}

\lref\ChandiaIX{
  O.~Chandia,
  ``A Note on the classical BRST symmetry of the pure spinor string in a curved background,''
JHEP {\bf 0607}, 019 (2006).
[hep-th/0604115].
}

\lref\BerkovitsQX{
  N.~Berkovits and O.~Chandia,
  ``Massive superstring vertex operator in D = 10 superspace,''
JHEP {\bf 0208}, 040 (2002).
[hep-th/0204121].
}

\lref\ChandiaHN{
  O.~Chandia and B.~C.~Vallilo,
  ``Conformal invariance of the pure spinor superstring in a curved background,''
JHEP {\bf 0404}, 041 (2004).
[hep-th/0401226].
}

\lref\BedoyaIC{
  O.~A.~Bedoya and O.~Chandia,
JHEP {\bf 0701}, 042 (2007).
[hep-th/0609161].
}

\lref\BerkovitsUE{
  N.~Berkovits and P.~S.~Howe,
  ``Ten-dimensional supergravity constraints from the pure spinor formalism for the superstring,''
Nucl.\ Phys.\ B {\bf 635}, 75 (2002).
[hep-th/0112160].
}

\lref\BakhmatovFPA{
  I.~Bakhmatov and N.~Berkovits,
  ``Pure Spinor $b$-ghost in a Super-Maxwell Background,''
JHEP {\bf 1311}, 214 (2013).
[arXiv:1310.3379 [hep-th]].
}

\newsec{ÊIntroduction}

Although the pure spinor formalism for the superstring \BerkovitsFE\ has several nice features such as manifest spacetime supersymmetry and a simple BRST operator, a complicated feature of the formalism which needs to be better understood is the pure spinor $b$ ghost.
Unlike in the bosonic string or Ramond-Neveu-Schwarz (RNS) string where the $b$ ghost is
a fundamental worldsheet variable, the $b$ ghost in the pure spinor formalism is a composite operator constructed from the other worldsheet variables. 

The BRST operator $Q$ in the pure spinor formalism is independent of the target-space
background and takes the simple form 
\eqn\brstpure{ Q = \int dz (\l^\a d_\a + \oh^\a r_\a)}
where $(\l^\a, d_\a, \oh^\a, r_\a)$ are fundamental worldsheet variables. To satisfy
the relation $\{Q, b\} = T$ where $T$ is the conformal stress tensor, $b$ was constructed in a flat background in
\BerkovitsBT\ as a complicated function of the worldsheet fields. This construction of the
pure spinor $b$ ghost was simplified in \BerkovitsPLA\ by introducing a fermionic vector field $\Gb_a$ which was related to the usual RNS fermionic vector field $\psi_a$ by ``dynamical twisting''.

An important question is how to generalize this construction of the pure spinor $b$ ghost in a curved superstring background. In \BakhmatovFPA, the pure spinor $b$ ghost was constructed in a super-Maxwell background of the open superstring, and in \ChandiaIMA, the pure spinor $b$ ghost was constructed in an N=1 supergravity background of the heterotic superstring.
The construction in \ChandiaIMA\  for a curved heterotic superstring background was quite complicated and some of the coefficients of the composite operator for the pure spinor $b$ ghost were not explicitly computed. 

In this paper, we will use the RNS-like fermionic vector field $\Gb_a$ of \BerkovitsPLA\ to
simplify the construction of the pure spinor $b$ ghost in a curved heterotic superstring background. When expressed in terms of $\Gb_a$, the composite operator for the pure spinor $b$ ghost in a curved heterotic background is a simple covariantization of the composite operator in a flat background that was found in \BerkovitsPLA.
In addition to simplifying the construction in a curved heterotic background, it is hoped
that this method involving $\Gb_a$ will also be useful for constructing the pure spinor $b$ ghost in an N=2 supergravity background of the Type II superstring.

In section 2, we review the pure spinor formalism of the heterotic superstring in a flat and curved target-superspace background. In section 3, we define the RNS-like variable $\Gb_a$
in a curved background. And in section 4, we use $\Gb_a$ to explicitly construct the pure spinor $b$ ghost in a curved heterotic superstring background. The appendix computes
the BRST transformations of $\Gb_a$ and the $b$ ghost in a flat background. 

\newsec{ÊReview of the Non-minimal Pure Spinor Formalism } 

In this section we review the non-minimal pure spinor formalism of the heterotic superstring in flat space \BerkovitsBT\ and in a curved background \ChandiaIMA.

\subsec{ Non-minimal pure spinor formalism in flat space}

The non-minimal pure spinor formalism in the heterotic superstring is constructed using the ten-dimensional $N=1$ superspace coordinates $(X^a , \t^\a)$ for $a=0$ to 9 and
$\a=1$ to 16,  the conjugate variable of $\t^\a$ which is called $p_\a$, a set of pure spinor variables $(\l^\a , \lh_\a , r_\a)$ together with their conjugate variables $(\o_\a , \oh^\a , s^\a)$, and the same 32 fermionic right-moving  variables $\xi^R$ for $R=1$ to 32 as in the RNS heterotic superstring formalism. The pure spinor variables are constrained to satisfy
\eqn\psc{ (\l \g^a \l) = (\lh \g^a \lh) = (\lh \g^a r) = 0 ,}
 where $(\g^a)_{\a\b}$ and $(\g^a)^{\a\b}$ are the symmetric gamma matrices which satisfy the Dirac algebra
\eqn\dirac{Ê(\g^a)_{\a\g} (\g^b)^{\g\b} + (\g^b)_{\a\g} (\g^a)^{\g\b} = 2 \eta^{ab} \d_\a^\b .}
Because of the pure spinor conditions \psc, the conjugate pure spinor variables are defined up to the gauge invariances
\eqn\gps{Ê\d\o_\a = (\l\g^a)_\a \L_{1a},\quad \d s^\a = (\g^a \lh) ^\a \L_{2a},\quad \d\oh^\a = (\g^a \lh)^\a \L_{3a} - (\g^a r) \L_{2a} ,}
where $\L_1, \L_2, \L_3$ are arbitrary gauge parameters. The action of the theory is quadratic in these variables and is given by
\eqn\frees{ÊS = S_0 + \int d^2 z ~  (\oh^\a \pb \lh_\a + s^\a \pb r_\a ) ,}
where
\eqn\sminF{ S_0 = \int d^2 z ~ (\ha \p X^a \pb X_a + p_\a \pb \t^\a + \o_\a \pb \l^\a + \xi^R \p \xi^R ) }
is the minimal pure spinor action.

The quantization of this system is performed by introducing a left-moving BRST charge given by
\eqn\brst{ÊQ = \oint dz (\l^\a d_\a + \oh^\a r_\a) ,}
where
\eqn\defd{Êd_\a = p_\a - \ha (\t\g^a)_\a \left( \p X_a + {1\over4}Ê(\t \g_a \p\t) \right) .}
This BRST charge is nilpotent because of the OPE
\eqn\dydz{ d_\a(y) d_\b(z) \to - {1\over(y-z)} \g^a_{\a\b} \Pi_a (z) }
where
\eqn\defpi{Ê\Pi_a = \p X_a + \ha (\t \g_a \p\t) .} 
The BRST transformations of the worldsheet fields of our system are
\eqn\qflat{ÊQ \Pi^a = - (\l \g^a \p \t),\quad Q \t^\a = \l^\a,\quad Q d_\a = \Pi^a (\g_a \l)_\a, \quad Q\l^\a = 0,\quad Q\o_\a = d_\a,} 
$$
Q \lh_\a = -r_\a,\quad Q \oh^\a = 0,\quad Q r_\a = 0,\quad Q s^\a = \oh^\a .$$
Note that the non-minimal sector in our system does not change the cohomology of the minimal sector and the action \frees\ can be written as
\eqn\sqsF{ S =S_0 + Q  \int d^2 z ~ s^\a \pb \lh_\a .}
 
The non-minimal pure spinor formalism does not contain the $(b,c)$ worldsheet reparameterization  ghosts. However, one can construct an operator $b$ satisfying the equation $Qb=T$ where $T$ is the world-sheet stress tensor of the action  \frees, and this operator is identified with the pure spinor $b$ ghost \BerkovitsBT. It was shown in \BerkovitsPLA\  that this $b$ ghost is simplified by introducing the RNS-like fermionic vector 
\eqn\gammaflat{ \Gb_a = - {1\over2(\l\lh)} (d \g_a \lh) -  {1\over8(\l\lh)^2} (r\g_{abc}\lh) N^{bc} }
where $N^{ab} = \ha (\l \g^{ab} \o)$ is the Lorentz current for the minimal pure spinors.

In terms of \gammaflat, the pure spinor $b$ ghost is 
\eqn\bflat{ b = -s^\a \p \lh_\a - \o_\a \p\t^\a + \Pi^a \Gb_a - {1\over4(\l\lh)} (\l \g^{ab} r) \Gb_a \Gb_b + {1\over2(\l\lh)} (\o \g_a \lh) (\l\g^a \p\t) .}
To verify the relation $Qb=T$, we first compute the action of $Q$ on $\Gb_a$ to be
\eqn\qGbflat{ÊQ \Gb_a = - {1\over2(\l\lh)} \Pi^b (\lh \g_a \g_b \l) - {1\over4(\l\lh)^2} (\l \g_{bc} r) (\lh \g^c \g_a \l) \Gb^b .}
The proof of \qGbflat\ and the verification of $Qb = T$ in a flat background are in the appendix. 

The purpose  of this paper is to find a $\Gb_a$ in a curved heterotic superstring background that satisfies an equation analogous to \qGbflat\ and to define the $b$ ghost as the covariantization of \bflat. In order to do this, we will need the BRST transformations corresponding to \qflat\ in a curved background. We now review the non-minimal pure spinor formalism in a curved heterotic superstring background.

\subsec{ Non-minimal pure spinor formalism in a curved background}

The minimal sector of the heterotic string in a curved background was constructed in \BerkovitsUE. The action has the form
\eqn\sminC{ÊS_0 = \int d^2z ~ (\ha \Pi_a \Pib^a + \ha \Pi^A \Pib^B B_{BA} + d_\a \Pib^\a + \o_\a \Nb \l^\a + \xi^R\N\xi^R +\ha\a' \Phi r ),}  
where $\Pi^A$ and $\Pib^A$ for $A=(a,\a)$ are defined from the background supervielbein $E_M{}^A$ and
the target superspace coordinates $Z^M$ as $\Pi^A = \p Z^M E_M{}^A$ and $\Pib^A = \pb Z^M E_M{}^A$, $B_{BA}$ is the graded-antisymmetric two-form superfield, $\Phi$ is the dilaton
superfield which couples to the two-dimensional worldsheet curvature $r$, 
\eqn\nab{\N\xi^R = \p\xi^R + T^{RS}_I \xi^S(\Pi^A A^I_A +d_\a W^{I\alpha}
+\ha  N_{ab} F^{I ab})}
where $T_I^{RS}$ are the SO(32) adjoint matrices for $I=1$ to 496 and $(A^I_M, W^{I \a}, F^{I ab})$ are the super-Yang-Mills gauge fields and field-strengths,
and
\eqn\covl{ \Nb \l^\a = \pb \l^\a + \l^\b \Pib^A \O_{A\b}{}^\a }
where the connection $\O_{A\b}{}^\a$ has the structure
\eqn\Ostr{Ê\O_{A\b}{}^\a =  \O_A \d_\b^\a + {1\over4}Ê\O_{Aab} (\g^{ab})_\b{}^\a . }
Here $\O_{Aab}$ is the usual Lorentz connection and $\O_A$ is a connection for scaling transformations introduced in \BerkovitsUE, and one can verify their coupling preserves the pure spinor gauge invariance of \gps.

The presence of the scale connection $\O_A$ in \sminC\ implies that the action is invariant not only under the usual local Lorentz transformations, but also under the local fermionic scale transformations 
\eqn\localscale{ \d \l^\a = \Lambda \l^\a, \quad \d \o_\a = -\Lambda \o_\a,\quad
\d d_\a = -\Lambda d_\a,}
$$\d\O_{\a\b}{}^\g = -(\p_\a \Lambda) \d_\b^\g - \L \O_{\a\b}{}^\g  ,\quad \d \O_{a\b}{}^\g = -(\p_a \L) \d_\b^\g ,$$ 
$$\d E_M{}^\a = \Lambda E_M{}^\alpha, \quad \d E_\a{}^M = -\Lambda E_\a{}^M.$$
So variables and superfields with raised tangent-space spinor indices carry charge $+1$ with respect to the fermionic scale transformations and variables and superfields with lowered tangent-space spinor indices
carry charge $-1$.

The minimal BRST charge is given by $Q_0 = \oint \l^\a d_\a$ and it was shown in \BerkovitsUE\ that nilpotency and holomorphicity of $Q_0$ forces the background to satisfy the equations of $N=1$ ten-dimensional supergravity. Nilpotency implies that 
\eqn\nilp{ \l^\a \l^\b T_{\a\b}{}^A = 0,\quad \l^\a \l^\b \l^\g R_{\a\b\g}{}^\d = 0,}
where $T_{\a\b}{}^A$ and $R_{\a\b\g}{}^\d$ are torsion and curvature components. And as shown in \BerkovitsUE, nilpotency and holomorphicity imply that the torsion components can 
be gauge-fixed to the form
\eqn\const{ T_{\a\b}{}^a = \g_{\a\b}^a, \quad T_{A \b}{}^\g =0, \quad T_{\a a}{}^b = 2 (\g_a{}^b)_\a{}^\b \O_\b.}
In addition, the absence of chiral \BerkovitsUE\ and conformal \ChandiaHN\ anomalies of the
worldsheet action implies that $\O_\a$ is related to the dilaton superfield $\Phi$ by
\eqn\dilaton{\Omega_\a = {1\over 4} D_\a \Phi.}


The BRST transformations of the minimal fields in \sminC\ were determined in \ChandiaIX\ to be
\eqn\QC{ÊQ\Pi^a = -\l^\b \O_{\b b}{}^a \Pi^b -\l^\a \Pi^b T_{b\a}{}^a,\quad  Q \Pi^\a =
-\l^\b\O_{\b \g}{}^\a\Pi^\g +  \N\l^\a,}
$$Q \l^\a = -\l^\b\O_{\b\g}{}^\a \l^\g,\quad Q\o_\a = \l^\b\O_{\b \a}{}^\g \o_\g + d_\a ,$$
$$ Q d_\a =  \l^\b \O_{\b\a}{}^\g d_\g + \Pi^a (\g_a\l)_\a + \l^\b \l^\g \o_\d R_{\a\b\g}{}^\d ,$$
where the first term in these transformations is a Lorentz and scale transformation proportional
to $\l^\b \O_{\b\g}{}^\a$.

For the non-minimal sector, it was noted in \ChandiaIMA\ that there is an effect of the background geometry on the BRST transformations of the non-minimal pure spinor fields.  Assuming that the minimal sector is unaffected by the non-minimal variables, a cohomological argument determined that $(\lh, \oh, r, s)$ transform in a curved background as 
\eqn\QQ
{ Q \lh_\a  = - r_\a + \l^\g\lh_\b (\O_{\g\a}{}^\b - {1\over 4}ÊT_{\g ab} (\g^{ab})_\a{}^\b),}
$$ Q \oh^\a = -\oh^\b\l^\g (\O_{\g\b}{}^\a - {1\over 4}ÊT_{\g ab} (\g^{ab})_\b{}^\a) ,$$
$$ Q s^\a = \oh^\a + s^\b\l^\g(\O_{\g\b}{}^\a - {1\over 4}ÊT_{\g ab} (\g^{ab})_\b{}^\a),$$
$$ Q r_\a = -\l^\g r_\b (\O_{\g\a}{}^\b - {1\over 4}ÊT_{\g ab} (\g^{ab})_\a{}^\b).$$
Note that the torsion $T_{\g ab}$ includes the Lorentz connection $\O_{\g ab}$, so the Lorentz part of the spin connection of \Ostr\ does not appear in these non-minimal BRST transformations.

To construct the non-minimal action in a curved background, the BRST-trivial term
\eqn\nonmincurved{ S_{non-min} = Q \int d^2 z ~ (s \Nb \lh + {1\over 4}\Pib^A T_{A ab}(s\g^{ab}\lh) ) }
will be added to the minimal action of \sminC\ where $\Nb\lh_\a = \pb\lh_\a - \lh_\b 
\Pib^A \O_{A \a}{}^\b$. This construction is analogous to the flat action of \sqsF, and although
the torsion term in \nonmincurved\ is not needed for covariance and was not included in \ChandiaIMA, it will simplify the
construction by decoupling the Lorentz connection $\O_{A ab}$ from the non-minimal action.
Using the BRST transformations of \QQ, one finds that 
\eqn\nonminr{S_{non-min} = \int d^2 z ~
( \oh^\a \Nb \lh_\a + s^\a \Nb r_\a + {1\over4} \Pib^A T_{A ab} ( \oh \g^{ab} \lh + s \g^{ab} r ) }
$$ + \l^\a \Pib^A R_{A\a} (s^\b \lh_\b) + {1\over4} \l^\a \Pib^A ( R_{A\a ab} - \N_{[A} T_{\a]ab} - T_{A\a}{}^c T_{cab} + T_{Ac[a} T_{b]\a}{}^c ) (s \g^{ab} \lh)) $$
\eqn\actionp{  = \int d^2z ~ (\oh^\a \Nb \lh_\a + s^\a \Nb r_\a + {1\over4} \Pib^A T_{A ab} ( \oh \g^{ab} \lh + s \g^{ab} r ) }
$$ + \l^\a \Pib^A R_{A\a} (s^\b \lh_\b) + {1\over4} \l^\a \Pib^d ( R_{d\a ab} - \N_{[d} T_{\a]ab} - T_{d\a}{}^c T_{cab} + T_{dc[a} T_{b]\a}{}^c ) (s \g^{ab} \lh)) $$
where we have used the Bianchi identity
\eqn\lpb{ R_{\b\a ab} - \N_{(\b} T_{\a) ab} - T_{(\b a}{}^c T_{\a) cb} - \g_{\b\a}^c T_{cab}=0 }
in the second line of \nonminr.

Using the Noether method, one can easily determine the BRST charge corresponding to the action of $S = S_0 + S_{non-min}$  to be
\eqn\brstnoether{ Q = \int dz (\l^\a d_\a + \oh^\a r_\a).}

\newsec{ Definition of $\Gb_a$ in Curved Background}

\subsec{ Simplified BRST transformations}

The first step in defining the curved background generalization of $\Gb_a$ of
\gammaflat\ is to define a new variable
\eqn\newD{Ê D_\a = d_\a + {1\over4} \l^\b T_{\b ab} (\g^{ab} \o)_\a - 3 (\l\O) \o_\a .}
In terms of $D_\a$, the BRST transformation of $\o_\a$ is given by
\eqn\qomega{ Q \o_\a = \l^\b\O_{\b \a}{}^\g \o_\g + D_\a + 3(\l\O)\o_\a - {1\over4} \l^\b T_{\b ab} (\g^{ab}\o)_\a .}
Furthermore, the BRST transformation of $D_\a$ is
\eqn\qD{ Q D_\a =   \l^\b \O_{\b\a}{}^\g D_\g+  \Pi^a (\g_a\l)_\a + 3 (\l\O) D_\a - {1\over4} \l^\b T_{\b ab} (\g^{ab} D)_\a }
$$ + \l^\b \l^\g \o_\d \left( R_{\a\b\g}{}^\d + {1\over4} (\g^{ab})_\a{}^\d \N_\g T_{\b ab} + {1\over16} (\g^{ab} \g^{cd})_\a{}^\d T_{\b ab} T_{\g cd} \right) $$
$$= \l^\b \O_{\b\a}{}^\g D_\g +  \Pi^a (\g_a\l)_\a + 3 (\l\O) D_\a - {1\over4} \l^\b T_{\b ab} (\g^{ab} D)_\a $$
where we have used that $Q(\l\O) = 0$ because $\O_\a$ is proportional to $\N_\a\Phi$.
To prove that the second line in \qD\ is zero, symmetrize in ${(\b\g)}$ and use the Bianchi identity $R_{(\a\b\g)}{}^\d=0$ and $\l^\b\l^\g R_{\b\g}=0$ to show that the second line is equal to
 $$ {1\over8} \l^\b \l^\g \o_\d \left( (\g^{ab})_\a{}^\d ( -R_{\b\g ab} + \N_{(\b} T_{\g) ab}Ê) + {1\over4} [\g^{ab} , \g^{cd}]_\a{}^\d T_{\b ab} T_{\g cd} \right) $$
 $$ = {1\over8} \l^\b \l^\g \o_\d \left(   \N_{(\b} T_{\g) ab}Ê - T_{a(\b}{}^c T_{\g)cb} - R_{\b\g ab} \right) (\g^{ab})_\a{}^\d = - {1\over8} \l^\b \l^\g \o_\d \g^c_{\b\g} T_{cab}(\g^{ab})_\a{}^\d  = 0 ,$$
where we used the Bianchi identity for $R_{[\b\g a]b}$. 

The BRST transformations of  \QQ, \qomega\ and \qD\ all involve a Lorentz and scale
transformation proportional to 
\eqn\propto{ - \l^\b \O_{\b\a}{}^\g +  {1\over4} \l^\b T_{\b ab} (\g^{ab})_\a{}^\g.}
It will be useful to define $ \tilde Q = Q - Q_{L+S}$ where $Q_{L+S}$ is this Lorentz and scale transformation, and
one finds that 
\eqn\QCC{Ê\tilde Q \Pi^a = -(\l \g^a \Pi),\quad \tilde Q \Pi^\a = \N \l^\a + {1\over4} \l^\b T_{\b ab} (\g^{ab} \Pi)^\a,\quad \tilde Q D_\a = \Pi^a (\g_a\l)_\a + 3 (\l \O) D_\a, }
$$ \tilde Q \l^\a = 5 (\l \O) \l^\a,\quad \tilde Q \o_\a = D_\a + 3 (\l \O) \o_\a ,$$
$$
\tilde Q \lh_\a = -r_\a,\quad \tilde Q \oh^\a = 0,\quad \tilde Q r_\a = 0,\quad \tilde Q s^\a = \oh^\a,$$
where we have used that 
\eqn\qtildel{\tilde Q\l^\a =  {1\over4}Ê\l^\b T_{\b ab} (\g^{ab} \l)^\a = \ha (\l \g_{ab} \O) (\g^{ab} \l)^\a = 5 (\l\O) \l^\a .}

\subsec{ Construction of $\Gb_a$}

In this subsection, it will be shown that 
\eqn\Gbc{ \Gb_a = -{1\over2 (\l\lh)} (D \g_a \lh) - {1\over8(\l\lh)^2} (r\g_{abc}\lh) N^{bc} }
satisfies the BRST transformation
\eqn\QGC{ÊÊ\tilde Q \Gb_a = - {1\over2(\l\lh)} \Pi^b (\lh \g_a \g_b \l) - {1\over4(\l\lh)^2} (\l \g_{bc} r) (\lh \g^c \g_a \l) \Gb^a - 2 (\l^\a \O_\a) \Gb_a }
where $\tilde Q$ is defined in \QCC. Comparing with the equations of 
\gammaflat\ and \qGbflat, one sees that \Gbc\ is constructed in a curved background by  replacing $d_\a$ in the flat-space construction with  $D_\a$ of \newD.

Since the BRST transformations of \QCC\ closely resemble the flat space BRST transformation
of \qflat, the only necessary step to proving \QGC\ is to show that the terms $(\l^\a \O_\a)$ which appear in \QCC\ sum up to $-2(\l^\a \O_\a) \Gb_\a$. From the first term in \Gbc, one obtains
\eqn\GDlh{Ê {1\over2(\l\lh)^2} (5(\l\lh)(\l\O)) (D \g_a \lh) -  {1\over2(\l\lh)} ( 3 (\l\O) D_\a) (\g_a \lh)^\a = 2(\l\O) {1\over2(\l\lh)} (D \g_a \lh ) ,} 
where the first term comes from the transformation of $(\l\lh)^{-1}$ and the second term comes from the transformation of $D_\a$.  And from the second term in \Gbc, one obtains
\eqn\GrN{Ê {1\over4(\l\lh)^3} (5(\l\lh)(\l\O)) (r \g_{abc} \lh) N^{bc}  
 +  {1\over8(\l\lh)^2} (r \g_{abc} \lh) ( 5 (\l \O) N^{bc} + 3 (\l\O) N^{bc} )}
$$= 2 (\l\O) {1\over8(\l\lh)^2}  (r \g_{abc} \lh) N^{bc} $$
where the first  term comes from the transformation of $(\l\lh)^{-1}$
and the second term comes from the transformation of $N^{bc}$. So we have proven \QGC.



\newsec{ Definition of $b$ Ghost in Curved Background}

In this section, we will use the dynamical twisting method of \BerkovitsPLA\ to simplify the construction
of the $b$ ghost in a curved heterotic background which was proposed in \ChandiaIMA. We will show that the $b$ ghost in a curved background can be defined in terms of the dynamically twisted RNS-like
variable \Gbc\ by simply covariantizing the flat-space expression of \bflat\ as 
\eqn\bgh{ b = -s^\a \N \lh_\a + {1\over 4}\Pi^A T_{A ab}(s\g^{ab}\lh)  - \o_\a \Pi^\a + \Pi^a \Gb_a - {1\over4(\l\lh)} (\l \g^{ab} r) \Gb_a \Gb_b + {1\over2(\l\lh)} (\o \g_a \lh) (\l\g^a \Pi) .}
As in the action of \nonmincurved, the torsion term in \bgh\ is not needed for covariance and was not included in \ChandiaIMA, but simplifies the construction by removing the dependence of the $b$ ghost on the Lorentz connection $\O_{A ab}$.

To prove that $Qb=T$ where $T$ is the stress-energy tensor of the heterotic string in a curved background, note that $S=S_0+ S_{non-min}$ of \sminC\ and \nonmincurved\ implies that
\eqn\curvedT{T = - \ha \Pi_a \Pi^a - d_\a \Pi^\a - \o_\a \N \l^\a + Q( -s^\a \N \lh_\a + {1\over 4}\Pi^A T_{A ab}(s\g^{ab}\lh) ).} 
So one needs to show that
\eqn\needcurved{Q b_{min} =- \ha \Pi_a \Pi^a - d_\a \Pi^\a - \o_\a \N \l^\a}
where 
\eqn\bmin{b_{min} = - \o_\a \Pi^\a + \Pi^a \Gb_a - {1\over4(\l\lh)} (\l \g^{ab} r) \Gb_a \Gb_b + {1\over2(\l\lh)} (\o \g_a \lh) (\l\g^a \Pi).}

Although $b_{min}$ is invariant under local Lorentz transformations, it transforms under the local scale transformation of \localscale\ as
\eqn\scaleb{\d b_{min} = -2\Lambda \Pi^a \Gb_a + 4\Lambda {1\over4(\l\lh)} (\l \g^{ab} r) \Gb_a \Gb_b }
where we have used that $\Gb_a$ of \Gbc\ transforms as $\d\Gb_a = -2\Lambda\Gb_a$.
Using the definition of $\tilde Q= Q - Q_{L+S}$ in \QCC, \needcurved\ is therefore implied if
\eqn\tildecurved{\tilde Q b_{min} +2(\l\O) \Pi^a\Gb_a - {(\l\O)\over(\l\lh)} (\l \g^{ab} r) \Gb_a \Gb_b  =- \ha \Pi_a \Pi^a - d_\a \Pi^\a - \o_\a \N \l^\a.}
Because of the similarity of \QCC\ with the flat space BRST transformations of \qflat\ and the result that $Q_{flat} b_{flat} = T_{flat}$, proving \tildecurved\ only requires showing that the various factors of $(\l^\a \O_\a)$ coming from  \QCC\ and \QGC\ cancel out in \tildecurved. 

The first term in \bmin\ contributes no factors of $(\l^\a \O_\a)$ and the second term in \bmin\
contributes $-2(\l\O) \Pi^a \Gb_a$ from the $\tilde Q$ variation of $\Gb_a$. The third term in
\bmin\ 
contributes 
\eqn\fourth{Ê{1\over4(\l\lh)^2} (5 (\l\lh) (\l\O) ) (\l \g^{ab} r) \Gb_a \Gb_b - {1\over4(\l\lh)} ( 5(\l\O)\l^\a) (\g^{ab} r)_\a \Gb_a \Gb_b +{(\l\O)\over(\l\lh)} (\l \g^{ab} r) \Gb_a \Gb_b}
$$  = {(\l\O)\over(\l\lh)} (\l \g^{ab} r) \Gb_a \Gb_b ,$$
where the first term comes from the variation of $(\l\lh)^{-1}$, the second term from the variation of $\l^\a$, and the third term from the variation of $\Gb_a\Gb_b$. Finally,
the fourth term in \bmin\ contributes
\eqn\fifth{Ê-{1\over2(\l\lh)^2} ( 5(\l\O) (\l\lh) ) (\o \g_a \lh) (\l \g^a \Pi) + {1\over2(\l\lh)} 3(\l\O) (\o \g_a \lh) (\l \g^a \Pi)}
$$ + {1\over2(\l\lh)} (\o \g_a \lh) 5 (\l\O) (\l \g^a \Pi) + {1\over2(\l\lh)} (\o \g_a \lh) (\l\g^a)_\a {1\over4}Ê\l^\b T_{\b bc} (\g^{bc}\Pi)^\a =0 $$
where the first term comes from the transformation of $(\l\lh)^{-1}$, the second term comes from the transformation of $\o_\a$, the third term comes from the transformation of $\l^\a$, and the last term comes from the transformation of $\Pi^\a$. To show that \fifth\ is zero, we have
used that 
\eqn\lafif{Ê{1\over4(\l\lh)} (\l \g_{bc} \O) (\l \g_a \g_{bc} 	\Pi) (\o \g^a \lh) = {1\over4(\l\lh)} (\l \g_{bc} \O) (\l ( [ \g_a , \g_{bc} ] + \g_{bc} \g_a ) \Pi) (\o \g^a \lh) }
$$ = {1\over(\l\lh)} (\l \g_{ac} \O) (\l \g^c \Pi) (\o \g^a \lh) +{1\over4(\l\lh)} (\l \g_{bc} \O) (\l  \g_{bc} \g_a \Pi) (\o \g^a \lh) $$
$$=Ê-  {3\over2(\l\lh)} (\l\O) (\l \g_a \Pi) (\o \g^a \lh) .$$
So we have proven that the $(\l\O)$ factors cancel out in \tildecurved, and therefore $Qb =T$.

\vskip 15pt

\vskip 10pt
{\bf Acknowledgements:}
NB would like to thank Sebastian Guttenberg and 
Luca Mazzucato for useful discussions.
The work of NB is partially financed by 
CNPq grant 300256/94-9
and FAPESP grants 2009/50639-2 and 2011/11973-4, and
the work of OC is partially financed by FONDECYT project 1120263.

\appendix{A}{Computations in a Flat Background}
 
In this appendix, we will prove the equation \qGbflat\ for $Q\Gb_a$ and the equation $Qb =T$ in a flat background.

Using  the BRST transformations of \qflat\ acting on $\Gb_a$ of \gammaflat, one finds
\eqn\QGF{ Q\Gb_a = - {1\over2(\l\lh)} \Pi^b (\lh \g_a \g_b \l) - {1\over2(\l\lh)^2} (\l r) (d\g_a \lh) - {1\over2(\l\lh)} (d\g_a r) } 
$$ - {1\over4(\l\lh)^3}(\l r) (r \g_{abc} \lh) N^{bc} - {1\over8(\l\lh)^2} (r \g_{abc} r) N^{bc} + {1\over16(\l\lh)} (r \g_{abc} \lh) (\l \g^{bc} d) .$$  
The term independent of $r$ agrees in \qGbflat\ and \QGF, and the term 
linear in $r$ in \QGF\ is 
\eqn\drone{ - {1\over2(\l\lh)^2} (\l r) (d\g_a\lh) -  {1\over2(\l\lh)}  (d\g_a r) +  {1\over16(\l\lh)^2} (r\g_a\g_{bc}\lh) (\l\g^{bc} d) }
$$=- {1\over2(\l\lh)^2} (\l r) (d\g_a\lh) + {1\over4(\l\lh)^2} (\l\g_b\g_a r) (d\g^b \lh) $$
$$= - {1\over4(\l\lh)^2}  (\l\g_a\g_b r) (d\g^b \lh) 
=  {1\over8(\l\lh)^3} (\l\g_{bc} r) (\lh \g^c \g_a \l) (d\g^b\lh) $$
which is the term linear in $r$ in \qGbflat. To go from the first line to the second line of \drone, 
we have used the identity
\eqn\ident{Ê \ha \d_\a^\d \d_\b^\g +  {1\over {16}} (\g^{bc})_\a{}^\g (\g_{bc})_\b{}^\d ={1\over 4}  \g^b_{\a\b} \g_b^{\g\d} -  {1\over 8}  \d_\a^\g \d_\b^\d  }
together with the pure spinor constraints of \psc.

Finally, the terms quadratic in $r$ in \QGF\ are
\eqn\rrone{Ê-{1\over8(\l\lh)^2} (r \g_{abc} r) N^{bc} - {1\over4(\l\lh)^3} (\l r) (r \g_{abc} \lh) N^{bc} }
$$=-{1\over8(\l\lh)^2} (r \g_{abc} r) N^{bc} -{1\over384(\l\lh)^3} (\l \g_{def} \g_{bc} \g_a \lh) (r\g^{def} r) N^{bc} $$
$$ =-{1\over8(\l\lh)^2} (r \g_{abc} r) N^{bc}  + {1\over16(\l\lh)^3} (\l\g_b\g_a \lh) (r\g^{bde} r) N_{de} $$
$$= -{1\over16(\l\lh)^3} (\l\g_a\g_b r) (r\g^{bde} \lh) N_{de},$$
which is the term quadratic in $r$ in \qGbflat.
To go from the first line to the second line of \rrone, we have used the identity $r_\a r_\b = {1\over96} \g^{def}_{\a\b} (r \g_{def} r)$. To go from the second line to the third line, we have
used that $(\l\g^b)_\a N_{bc} = \half J (\g^c \l)_\a$ where $J =- \lambda^\a\o_\a$ and that all terms proportional to $J$ vanish using the pure spinor constraints of \psc.
And to go from the third line to the fourth line, we have used that 
$(\g_b r)^\a (\g^b \lh)^\b = - (\g_b \lh)^\a (\g^b r)^\b .$ So we have proven that $\Gb_a$ satisfies equation \qGbflat\ in a flat background.. 

We now verify that $Qb = T$ in flat space. Applying $Q$ of \qflat\ to \bflat, we obtain
\eqn\Qonb{ Q b = T + (\l \g^a \Pi) \left(  {1\over8(\l\lh)^2} (r \g_{abc} \lh) N^{bc} - {1\over2(\l\lh)^2} (\l r) (\o \g_a \lh) + {1\over2(\l\lh)}  (\o \g_a r) \right) }
$$ - {1\over4(\l\lh)^2} \Gb^a \Gb^b \left( (\l r) (\l \g_{ab} r) + {1\over2(\l\lh)} (\l \g_a{}^c r) ( \l \g_{bd} r) (\lh \g^d \g_c \l) \right) .$$ 
Using the identity
\eqn\idllor{Ê(\l\lh) (\o \g_a r) - (\l r) (\o \g_a \lh) = -{1\over8} (\lh \g_{abc} r) N^{bc} - {1\over48} (\l \g_{abcd} \o) (r \g^{bcd} \lh) ,}
we obtain that the second term in \Qonb\ is equal to
\eqn\termun{ {1\over16(\l\lh)} (\l\g^a\Pi) \left( (r \g_{abc} \lh) N^{bc} + {1\over6}Ê(\l \g_{abcd} \o) (r \g^{bcd} \lh) \right) ,}
which can be seen to vanish using $(\l\g^a)_\a (\l \g_a)_\b=0$.
Finally, the term proportional to $\Gb^a \Gb^b$ in \Qonb\ vanishes using the identities $(\l\g^a)_\a (\l \g_a)_\b = (\l r) (\l r) = 0$ .

 \listrefs
 
\end